\documentclass[12pt]{article}
\usepackage{amsmath}
\begin{document}
\title     {Reply to Comments on the article ``M.~O.~Katanaev, Complete
separation of variables in the geodesic Hamilton--Jacobi equation in four
dimensions, Physica Scripta (2023), 98, 104001'' by V.~V.~Obukhov,
K.~E.~Osetrin, and A.~V.~Shapovalov}

\author    {M. O. Katanaev
            \thanks{E-mail: katanaev@mi-ras.ru}\\ \\
            \sl Steklov Mathematical Institute,\\
            \sl 119991, Moscow, ul. Gubkina, 8\\
            and\\
            Lobachevsky Institute of Mathematics and Mechanics,\\
            Kazan Federal University, Kazan, Russia}
\maketitle
It is good that the authors of the Comments did not find a mistake in my papers
\cite{Katana23A,Katana23B}. The main criticism of the Comments is that my
papers ``do not contain any new results'' as compared to the paper
\cite{Shapov80}. I knew this paper when I wrote mine and even cited it in
\cite{Katana23A}.

Let me comment the article \cite{Shapov80}. V.~Shapovalov formulated the main
result as:

``Theorem 1. In a St\"ackel space, and only in such, there exists a
(priviledged) coordinate system $u$ wherein the metric tensor has the form
\begin{equation*}
\begin{split}
  &\overset\circ g{}^{ir}=f^{-1}_{\sigma\nu}(u)h^{ir}_\nu\left(
  \overset\nu u\right),
\\
  h^{ir}_\nu\equiv\delta^i_{i_\nu}\delta^r_{r_\nu}h^{i_\nu r_\nu}&
  +\big(\delta^i_{i_0}
  \delta^r_{r_\nu}+\delta^i_{r_\nu}\delta^r_{i_0}\big)h^{i_0 r_\nu}
  +\delta^i_{i_0}\delta^r_{r_0}h^{i_0 r_0}_\nu.\qquad\text{''}
\end{split}
\end{equation*}

The matrix $f$ in this ``theorem'' is described but there is no definition of
$h^{i_\nu r_\nu}$, $h^{i_0 r_\nu}$, and $h^{i_0 r_0}_\nu$ in the entire paper.
Therefore it is not a theorem, because there is no statement. Moreover, on page
793, V.~Shapovalov wrote: "Since Theorem 1 has no bearing on this question and
since its proof is very cumbersome, we omit it". As far as I know, the ``proof''
was never published.

Therefore I cannot criticize the ``result'' of V.~Shapovalov, because there is
neither theorem nor proof. In addition, the paper does not contain any example.
It is clear that ``Theorem 1'' by V.~Shapovalov is not related to
theorems proved in \cite{Katana23A}. For example, the St\"ackel matrix, which
is crucial for the whole construction, denoted by $f$ in \cite{Shapov80}, depends
only on coordinates. The analog of this matrix denoted by $b$ in Eq.(94)
\cite{Katana23A} depends also explicitly and nontrivially on parameters entering the
complete integrals of the Hamilton--Jacobi equation. The difference is evident.

In papers \cite{Katana23A,Katana23B}, I listed ten classes of separable metrics
in four dimensions. As far as I know, this is a new result. This list is not
mentioned in any of the papers cited in the Comments. In particular, the authors
of the Comments wrote: ``These theorems have been proved half a century ago by
Vladimir N.~Shapovalov. This also applies to the complete list
\cite{Shapov80,Shapov78A} of metrics, integrals of motion and complete integrals
of the free Hamilton--Jacobi equations in the privileged (separable) coordinate
systems for an arbitrary 4-dimensional St\"ackel space $V_4$ given in the
article \cite{Katana23B}". It is easily checked that there are no lists of
metrics, integrals of motion and complete integrals in
\cite{Shapov80,Shapov78A}.

Moreover, V.~V.~Obukhov and K.~E.~Osetrin published two books on the subject in
Russian \cite{Obukho06,ObuOse07}. On page 22 of \cite{ObuOse07}, it is written:
``It can be easily shown that there are only 7 types of St\"ackel spaces of
Lorentz signature $(+,-,-,-)$ -- three isotropic: (3.1), (2.1), and (1.1) and
four nonisotropic: (3.0), (2.0), (1,0), and (0.0)'' (the translation is mine).
Afterwards, the separable metrics are listed. We see that several separable
metrics are missed. This statement shows clearly that the list of separable
metrics in four dimensions was not known to the authors of the Comments at least
in 2007.

{\bf Acknowledgments}

This paper has been supported by the Kazan Federal University Strategic Academic
Leadership Program.

Many thanks to V.~V.~Obukhov, K.~E.~Osetrin, and
A.~V.~Shapovalov for the interest in my papers.

\end{document}